\documentclass[12pt,preprint]{aastex}

\shortauthors{Mazzei et al.}
\shorttitle{THE ($NUV-r$) vs. $M_r$ PLANE AS A TRACER OF ETGs EVOLUTION IN GROUPS}

\begin{document}

\newcommand{\virgolette}[1]{``#1''}

\title{THE ($NUV-r$) vs. $M_r$ AS A  TRACER OF EARLY-TYPE GALAXIES  EVOLUTION IN 
USGC U376 AND LGG~225 GROUPS}

\author{
Paola Mazzei\altaffilmark{1},
Antonietta Marino\altaffilmark{2},
Roberto Rampazzo\altaffilmark{1}
}
\altaffiltext{1}{INAF - Osservatorio Astronomico di Padova, Vicolo dell'Osservatorio, 5, 35122, Padova, Italy; 
 paola.mazzei@oapd.inaf.it}
\altaffiltext{2}{Universit\`a di Padova, Dipartimento di Fisica e Astronomia G. Galilei, Vicolo dell'Osservatorio, 3, 35122, Padova, Italy} 

\begin{abstract}

With the aim of tracing back the evolution of  galaxies in nearby groups we use 
smooth particle hydrodynamical (SPH) simulations with chemo-photometric implementation. 
Here we focus on the evolution of the early-type members (Es and  S0s, ETGs hereafter) in two groups, 
USGC U376 and LGG~225, both in the Leo  cloud. We use the near-UV (NUV)-optical rest-frame  $(NUV-r)$  versus
$M_r$ color magnitude diagram to follow their evolution, from 
the blue cloud (BC) to the red sequence (RS), through the  green valley (GV). 
ETGs brighter than $M_r=-21$\,mag are older than 
13 Gyr and spend up to 10 Gyr of their overall evolutionary time in the BC before they reach the RS migrating through the GV. 
Fainter ETGs are younger,  of $\approx$ 2\,Gyr  on average, and evolve for
about 7-8\,Gyr along the BC. The turn-off occurs at $z\approx 0.3-0.4$.
Therefore these ETGs spend up to 3-5\,Gyr  crossing the GV;  UGC~06324, the faintest ETG in the sample, still is
in the GV. The mechanism driving their evolution is gravitational, due to merging
and/or interaction.  Our SPH simulations 
suggest that ETG members of these groups evolved toward the RS before and 
during the group collapse phase. This result is consistent with the dynamical analysis of both groups showing that they
 are not yet virialized.
\end{abstract}

\smallskip

\section{Introduction}

The color distribution of galaxies in the local universe,
is nearly bimodal  and relates to galaxy morphology \citep[e.g.][]{Strateva01, Balogh04}.
In the  color-magnitude diagram (CMD),  quiescent early-type galaxies (ETGs)  populate the red sequence (RS) and  late-types, 
with active star formation (SF), populate the blue cloud \citep[BC; e.g.][]{Baldry04}.
This bimodality is ubiquitous, extending from the field, to groups
and clusters \citep[e.g.][]{Lewis02}.
The physical origin of this color distribution is still under debate. However,
there is  strong evidence that the two distinct populations are the result of 
transformations driven by the  environment.
Galaxy evolution from the BC to the RS, i.e., from star-forming to quiescent galaxies,
occurs via a transition that leads galaxies in an intermediate zone of the CMD, the green valley \citep[GV;][]{Martin07}.  
Since ultraviolet (UV) bands are  excellent tracers of the recent  SF, 
the red and blue color sequences are especially well separated in the far-UV (FUV)-optical and near-UV (NUV)-optical
CMDs \citep[e.g.][]{Schaw07,Wyderetal07}. In particular, FUV studies focused 
on the galaxies located in the GV, the paradigm of the transition phase. 
 FUV, with {\it Galaxy Evolution  Explorer}   \citep[GALEX;][]{Martin05}, and/or H$\alpha$ images 
revealed  signatures of ongoing SF in the form of inner and/or outer blue ring/arm-like 
structures in some  ETGs \citep{Salim10, Marino11b}. \citet[][and references therein]{Salimetal12}  
showed a wide collection of these kind of galaxies
at redshift $\le$0.12.   
Signatures of ongoing ($\approx$ 9$^{+4}_{-3}$\%) or recent ($\approx$47$^{+8}_{-7}$\%) 
SF  are also found in the nuclear regions  of nearby ETGs \citep[e.g.][]{Rampazzo13}.
Such ETGs have been proposed as possible examples of the ongoing transition 
\citep{Marino11a,Salimetal12,Fangetal12}.  \\
\indent
Morphology, colors, and the star formation rate (SFR), primarily 
depend on small-scale ($<$\,1\,Mpc) environment \citep{Hogg04, Kauff04, Bla05, Ball08, Wetzeletal12}.
This result  has been extended through  the use of galaxy group catalogs, showing that colors and SF
history most directly depend on the properties of their host dark matter (DM)  halo \citep{BlaeBe07, 
Wiletal10, Tietal12}, in agreement with  results of smooth particle hydrodynamical  SPH  simulations by \citet[][MC03 hereafter]{paola1}. 
In this context, the investigation of  the evolution of group members in the
nearby universe acquires a great cosmological interest because more than 
half of galaxies reside in  such environments.  Furthermore, since the 
velocity dispersion of galaxies  is significantly lower in groups than in clusters, 
the merger probability and the effects of interaction  on galaxy evolution
are much higher. Consequently, groups  provide a zoom-in on phenomena  
driving the galaxy morphological and SF evolution
before galaxies  fall into denser environments \citep[e.g][]{Wilman09,Just10}. 
In \citet{Ant13}, we investigated 
the UV--optical CMD of  the members of three nearby galaxy groups.
Here, we present SPH simulations with chemo-photometric 
implementation based on evolutionary population synthesis (EPS) models, 
aiming at studying the evolution of  all the ETGs in two of these
groups, namely USGC U376  \citep[U376 hereafter,][]{Ramella02}, and LGG~225
\citep{Garcia93}. Our SPH simulations allow us  to derive dynamical and
morphological information concurrently with the spectral energy distribution (SED) at each evolutionary
time (\virgolette{snapshot}, hereafter). Moreover, these simulations allow us to trace the evolutionary path of  each ETG,  from 
the BC to the RS through the  GV, in the ($NUV-r$) versus $M_r$  plane.\\
\indent
Simulations of groups, focusing on galaxy morphological transformation, have been 
approached with different techniques in the literature (e.g., \citet{Kawata08, Bekki11,Villalobos12}).
\citet{Kawata08} used a cosmological chemo-dynamical simulation to study 
how the group environment affects the SF properties of a disk galaxy.    
 \citet{Bekki11}, starting from already-formed galaxies, showed that
spirals (S) in groups can be strongly influenced by repetitive, slow encounters
so that galaxies with thin disks can be transformed into thick disks and gas-poor S0s.
\citet{Villalobos12} studied the  evolution of disk galaxies,
composed of a stellar disk embedded in a DM halo, within a group environment. 
The group is modeled using an N-body DM halo following a \citet{NFW97} density profile and including
a spherically symmetric stellar component at its center, to account for the central galaxy.\\
The novelty of our approach is that we explore the merger/interaction scenario starting from collapsing
triaxial systems composed of DM and gas in different proportions and
combine the SPH code with a chemo-dynamical code based on  
EPS models. The SED we derive accounts for chemical  evolution, stellar  emission, internal extinction, and 
re-emission by dust in a self-consistent way, as descibed in previous works 
\citep[][and references therein]{marilena, marilena12}. Each simulation
self-consistently provides morphological, dynamic, and chemo-photometric
evolution.
 Based on this powerful tool, we aim to trace back the evolution of ETGs in nearby groups. \\
\indent
The plan of the paper is as follows.  Section~2 summarizes the main
prescriptions of our SPH simulations, fully described in previous works
\citep[and references therein]{paolanew} and provides the initial
conditions of galaxy encounters that best-fit the global properties of selected
ETGs. Section~3 details the results of individual galaxies, in terms of
gas accretion history drawn by the simulation that provides the best-fit of their absolute magnitude, SED,
and morphology.  Section 4 presents and discusses the  evolutionary predictions in the 
($NUV-r$) versus $M_r$, CMD. Conclusions are drawn in Section~5.\\
\indent
Throughout the paper we assume $H_0$=70\,km\,s$^{-1}$\,Mpc$^{-1}$,
 $\Omega_{\Lambda}$=0.73, and $\Omega_{matter}$=0.27, as in {\tt HYPERLEDA}\footnote{http://leda.univ-lyon.fr1} database \citep{Paturel03}. 

\section{ETGs in U376 and LGG~225: data and simulations.}

 Table~\ref{table1} summarizes photometric and structural parameters of ETGs 
in both groups.
For each group (column 1) and galaxy (Column~2), the morphological type (Column~3) is  from \citet[][their Table 1]{Ant13};
 the  cosmological corrected scale  (Column~4) is derived from {\tt NED}\footnote{NASA/IPAC Extragalactic Database http://ned.ipac.caltech.edu};  
the apparent diameter corrected for Galactic extinction and inclination, 
$D_{25}$  (Column~5), is from  {\tt HYPERLEDA}; the effective radii, $r_{e}$ (Column 6), from {\tt RC3} catalog \citep{deV91} 
or from labeled references;  the $B$-band apparent magnitude corrected for Galactic extinction  (Column~7) is
from  {\tt HYPERLEDA}. The   $D_{25}$ and $r_e$ values are reported in kpc using  cosmology--corrected scales in the same Table. 
The total absolute 
magnitude in the  $B$-band (Column~8), is taken from {\tt HYPERLEDA} (left) and derived from the 
{\tt NED} distance modulus (right). We point out that the discrepancy of the above values is due
 to  the different corrections applied.  The left value  of M$_B$ accounts for the Virgo infall only, 
 while the right values relates to the 3\,K cosmic  microwave backgroud (CMB).
 We adopt the range of the values  above to constrain our simulations (Section 3).\\
\indent
The measured UV and optical total magnitudes are from   \citet[][their Table~6]{Ant13} 
for U376 and \citet[][their Tables 3 and 4]{Ant10} for LGG~225. This set has been extended in wavelength 
using Two Micron All Sky Survey (2MASS) and far-infrared (FIR) data from {\tt NED}  plus the {\tt AKARI/FIS} catalog. 
The observed SEDs of almost ETGs extend from the far-UV to 160\,$\mu$m, i.e., 
over about three orders  of magnitude in wavelength (Figure \ref{figure1}). \\
 These SEDs will be compared with those derived from our chemo-photometric simulations.\\
\indent
The general prescriptions of  SPH simulations and the grid of the impact parameters explored, are reported in several 
previous papers \citep{paolanew, Dani12,Trinetal12,marilena12}. 
All of our simulations of galaxy formation and evolution start from collapsing
triaxial systems composed of DM and gas in different proportions and different total masses. 
In more detail, each system is built up with a spin parameter,
$\lambda$, given by $|{\bf {J}}||E|^{0.5}/(GM^{0.5})$, where $E$ is the
total energy, $J$ is the total angular momentum and $G$ is the gravitational
constant;  $\lambda$ is equal to 0.06 and aligned with the shorter
principal axis of the DM halo. The triaxiality ratio of the DM halo,
$\tau=(a^2-b^2)/(a^2-c^2)$,  is 0.84 where $a>b>c$ \citep{war92}.
All the simulations  include self-gravity of gas, stars and DM, 
radiative cooling, hydrodynamical pressure, shock heating, 
 viscosity, SF,  feedback both from 
evolved stars and type II SNe, and  chemical enrichment.\\
\indent
Simulations provide the synthetic SED, based on EPS models, at each  evolutionary stage, i.e. at each snapshot. 
The time step between each snapshot is 0.037\,Gyr.
The SED accounts for chemical  evolution, stellar  emission, internal extinction, and 
re-emission by dust in a self-consistent way 
 \citep[][and references
therein]{marilena,marilena12}.
This extends over four orders of magnitude in wavelength at least,
i.e., from 0.06 to 1000 $\mu$m. Each simulation self-consistently 
 provides morphological, dynamic, and chemo-photometric evolution. 
 The initial mass function (IMF) is of Salpeter type with an
upper mass limit of 100\,$M_\odot$ and a lower mass limit of 0.01$\,M_\odot$
\citep{Salp55}, as in  \citet{CM99} and MC03.\\
\indent
All the model parameters  had been tuned in previously cited  papers  that
analysed the evolution of isolated collapsing triaxial halos, initially composed of
DM and gas.  In those papers the role of 
the initial spin  of the halos, their total mass and gas fraction,
as well as   different IMFs, particle resolutions, SF efficiencies, and values of the feedback parameter were
all examined.  The integrated properties of simulated galaxies,  stopped at 
15\,Gyr, i.e., their colors, absolute magnitudes, metallicities, and
mass to luminosity ratios,   had been successfully compared with those of local galaxies 
 (\citet[][their Fig. 17]{CM99}, \citet[][their Fig. 8]{paolaap,paolaa}). 
In particular, from  our IMF choice a slightly higher SFR  arises compared with the other possibilities examined; 
this allows for the lowest feedback strength and for the expected
rotational support when disk galaxies are formed \citep[][MC03]{CM99}. As pointed
out by \citet{K12}, this  slope is almost the same as the universal mass
function that links  the IMF of galaxies and stars to those of brown dwarfs,
planets and small bodies (meteoroids, asteroids) \citep{BH07}.\\
\indent
 The particle resolution is enhanced here to
6--8$\times$10$^4$ instead of  1--2$\times$10$^4$ as in MC03, so there are
3--4$\times$10$^4$ particle of gas and 3--4$\times$10$^4$ of DM at the
beginning in each new simulation.  
The gravitational softening is 1, 0.5, and 0.05\,kpc for DM and gas
and star particles, respectively.\\
\indent
From the grid of physically motivated SPH simulations, we isolate
those  simultaneously  best fitting the global properties of selected ETGs, i.e. their absolute magnitude,  
integrated SED, and  current morphology. 
 ETGs in our groups cannot be
matched by a single collapsing triaxial halo within the same framework and 
set of parameters. Mergers and/or interactions are,
indeed, strongly favored in groups, as discussed in Section 1.\\
We point the reader to the paper by \citet{paolanew} where our approach
to  match  the photometric, structural (e.g., disk versus bulge), and kinematical (gas versus stars kinemtics)  of two  S0 galaxies, 
NGC~3626 in the U376 group and NGC~1533
in the Dorado group \citep{Firth06},  has been discussed in great detail.\\
We put here  a further constraint to our simulations,  accounting for the dynamical analysis of each group \citep[][see also Appendix]{Antonew}.
On this basis, the sum of masses of simulated 
systems at their derived current ages, have to be consistent, i.e., they cannot exceed 
the dynamical mass of each group, as given in Table \ref{tableA1} (see the Appendix).\\
\indent
Table~\ref{table2} provides the initial conditions of the encounter of the two halos 
able to match the global properties of ETGs in Table \ref{table1}, as will  be discussed in the next section.
In particular, Column 3 of Table \ref{table2} gives the total initial number of particles (the number of DM particles 
is the same as gas particles), the length of the semi-major axis of the primary halo (Column 4), 
the pericentric separation of the halos in unit of their semi-major axis (Column 5), the  distance 
of the halos centers of mass from the center of mass of the global system (Columns 6 and 7),
the velocity moduli of the halo centers in the same frame (Columns 8 and 9), and
the  total mass (Column 10).  The initial gas mass resolution, $m_{gas}$, 
is  between 1.33$\times$10$^7$\,$M_{\odot}$ and 5$\times$10$^5$\,$M_{\odot}$.
The SF efficiency is 0.4. The value of this parameter  is set to
account for our previous works on isolated collapsing systems where
the effect of different values (i.e. 0.2 in \citet{CM99} and 0.04 in MC03) were
analyzed and discussed. MC03 concluded  that the SFR neither depends on mass gas resolution
nor on the SF efficiency being driven by the total amount of the gas. As far as
the stellar mass resolution is concerned, we point out that the SF
proceeds, if allowed by the onset of two conditions at the same time
\citep[see also][]{Val2002}, no more than 10 times in a single cloud. This
implies that the (gas) stellar mass resolution ranges by a factor of 10, from
(0.6) 0.4$\times\,m_{gas}$ to (0.06) 0.04$\times\,m_{gas}$. It turns out
that a value of efficiency of about 0.5 gives rise to a conversion of
gas-into-stars efficiency of about few percent per free-fall time (for
the chosen density threshold it is about 20 Myr), when a strong and
efficient thermal stellar feedback is present  \citep{GonSam13}. These values are close to
the value estimated for the giant molecular clouds in the Milky Way by \citet{KrueTan07}.\\
\indent
The SFR, which drives the evolution of the global
properties of our simulated galaxy, converges when the initial particle
number is above 10$^4$   \citep[MC03 for a discussion, their Fig.1;][]{Chrietal2010, Chrietal2012}.\\
\indent
The final number of particles doubles at least the initial number in Table \ref{table2}.\\
The gas-to-DM fraction is 0.11 in all the  simulations presented here.
 This value is very similar to the value of 0.13 found by \citet{Gonzetal2013}  
by analyzing a large sample of clusters with different total masses.

\section{Comparing simulations and data}

\indent
Among the large set of simulations we performed,  for each galaxy  we selected the 
snapshot that simultaneously  best reproduces the current $B$-band total absolute magnitude, 
the observed SED, and the galaxy morphology.  This snapshot provides,  in turn,  
several quantities, e.g., the current galaxy age. Furthermore, the corresponding
simulation self-consistently  provides  the evolutionary path of all the properties of the system components \citep[see for details][and references therein]{paolanew}. \\
\indent
In Figure \ref{figure1}, we show  the match between the synthetic SED
from our chemo-photometric SPH simulations and the multi-$\lambda$ observations for each galaxy.
The corresponding gas accretion histories, i.e., the evolution of the
gas mass inside a specific radius, is shown in Figure \ref{figure2}. The observed and simulated $r$-band maps,  on the same scale and 
spatial resolution (5\arcsec), are compared in Figure \ref{figure3} (right and middle columns).\\
\indent
Most of the  analyzed  ETGs are well modeled by a single major
merger of two halos (with a mass ratio of 1:1 or 1:2). Only UGC~06324 in U376 and NGC~3457 in LGG~225 are described by a galaxy 
encounter.\\
In Table \ref{table3}, we summarize the main results obtained from chemo-photometric SPH simulations.
In particular, for each group and ETG (Columns 1 and  2 respectively), we report 
 within the respective  $R_{25}$ ($D_{25}$/2, see Table \ref{table1}),
the galaxy age (Column 3),  the average age of  the stellar 
population weighted  by $B$-band luminosity, $t_{pop}$, (Column 4), the average stellar  metallicity (Column 5), the total mass (Column 6), 
the $M/L$ ratio (Column 7), the DM and  gas mass fraction  (Columns 8 and 9) and the $B$-band  total absolute magnitude (Column 10).\\
\indent
We point out that the total mass of all simulated systems in  U376   does not exceed the virial mass of the group itself 
(Table  \ref{tableA1}) just from the beginning (Table \ref{table2}), fulfilling the condition set in Section 2.
In the case of LGG~225, the total initial mass of our simulations exceeds 30\%  of the virial mass of this group.
However, the total  mass of simulated systems inside the virial radius ($\simeq$1.5 times the harmonic radius in Table \ref{tableA1}) at 
their current age  fulfills the condition set, as discussed in Section 3.2. \\
 In the following subsections we provide  details of the results for the individual ETGs.

\subsection{U376: Individual Notes of its ETGs}

\begin{figure*}
\centering
\includegraphics[width=10cm]{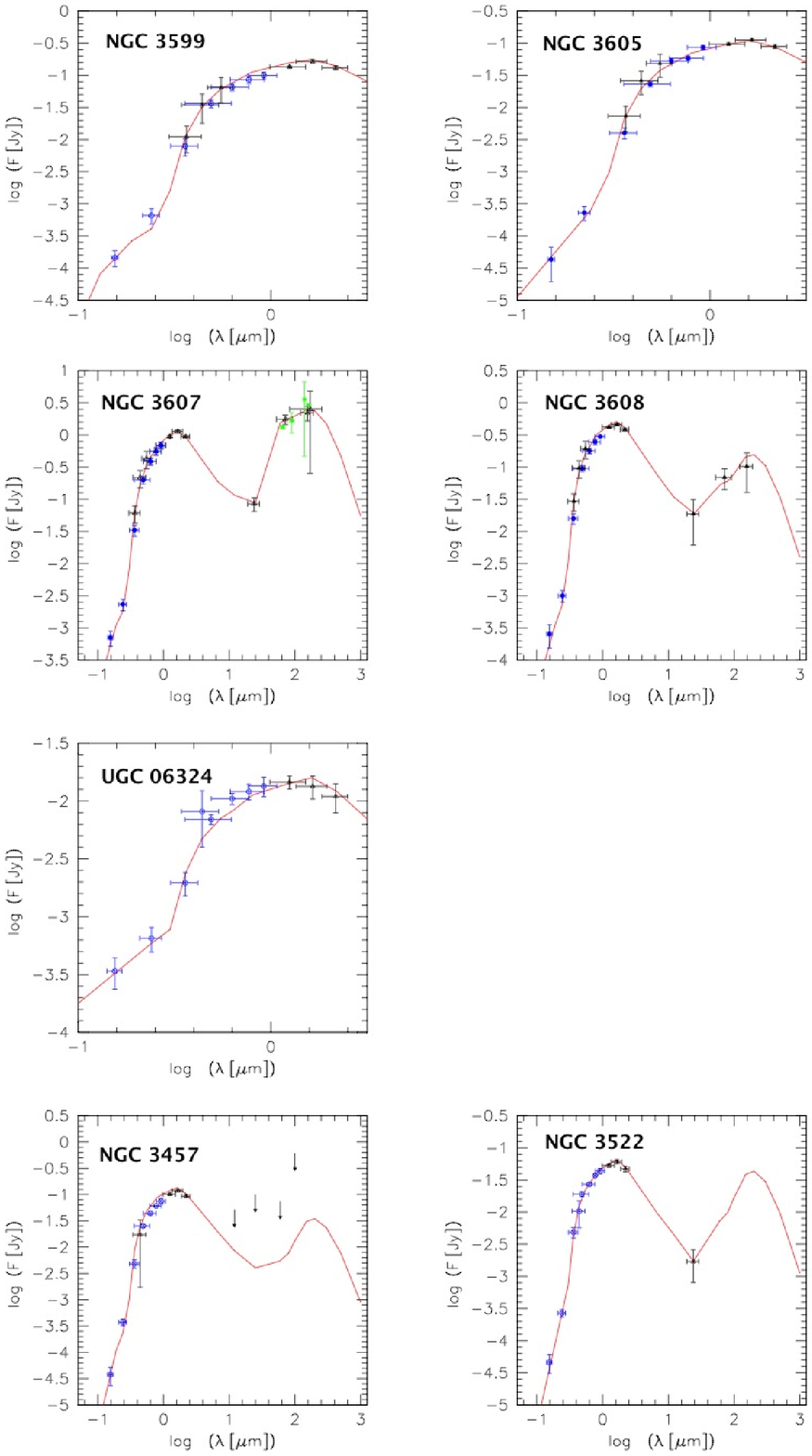}
\caption{Blue circles are 
the total fluxes  from   \citet[][their Table~6]{Ant13} 
for U376 and \citet[][their Tables 3 and 4]{Ant10} for LGG~225 (bottom panels); (black) triangles are 
from {\tt NED} and green asterisks are
from the {\tt AKARI/FIS} Bright Source Catalog \citep{Yamaetal09}. The error bars account 
for band width and 3\,$\sigma$ uncertainties of the fluxes; 
the (red) solid lines show our  predictions; arrows (left bottom panel) show $IRAS$ upper limits.} 
 \label{figure1}
\end{figure*}

\begin{figure*}
\centering
\includegraphics[width=10cm]{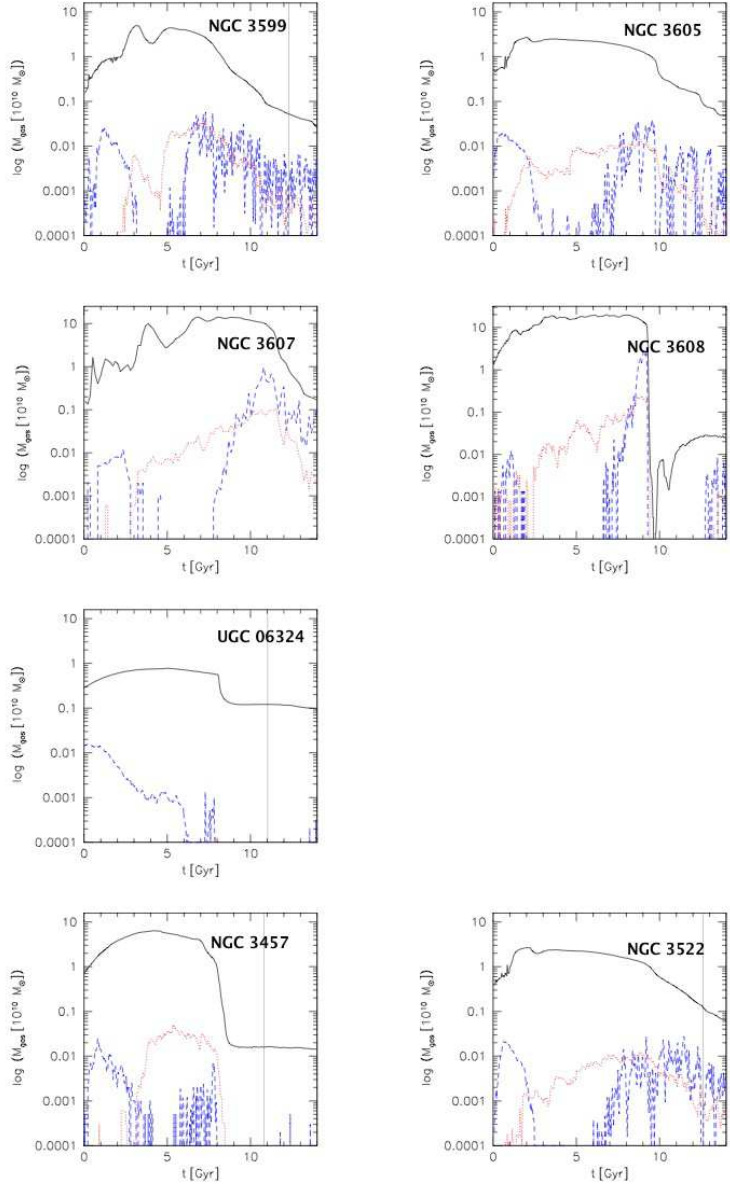}
\caption{
  Gas accretion history, i.e. the evolution of the gas mass inside a radius of 50\,kpc 
on the $B$-band luminous center of each galaxy, solid line, from SPH simulations. The (blue) dashed  and 
 the (red) dotted lines correspond to the gas with 
 temperature $\le 10^4$\,K and $\ge 10^6$\,K, respectively. The vertical lines mark the current  
age of galaxies younger than 13.8\,Gyr (Table \ref{table3}). }
 \label{figure2}
\end{figure*}

\begin{figure*}
\centering
\includegraphics[width=9cm]{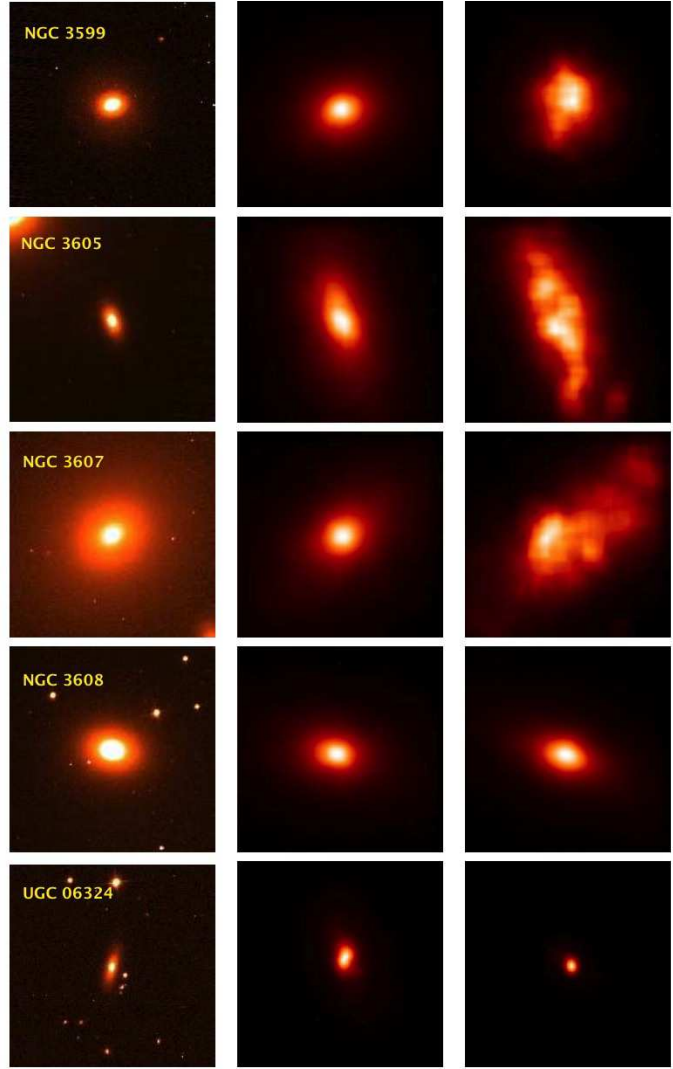}
\caption{
  SDSS $r$-band images of ETGs in U376 (left) and our simulated maps  (middle) on the same band, spatial scale, resolution (5\arcsec)
and density contrast (3) as the observed ones at the current galaxy age and 
 at the age corresponding to the brightest M$_r$(AB) magnitude  in Figure \ref{figure5} (right).
Maps are normalized to their total fluxes.}
\label{figure3}
\end{figure*}

\medskip\noindent
{\it{\object{NGC~3599}}}~~~~~
The global properties of this lenticular galaxy are well matched by a major  merger with a mass ratio of 1:1.
The spins of the merging systems are equal ($\lambda$=0.06, MC03), parallel,  both are aligned with the shorter of their principal axes, and perpendicular to the orbital plane (direct encounter). 
The  $B$-band total absolute magnitude,  $M_B=-18.4$\,mag,   and the SED  of the selected snapshot match  well with the observed values 
(Table~1 and Figure \ref{figure1}).
 The Sloan Digital Sky Survey (SDSS) $r$-band
morphology is compared with the simulated one on the same scale, 5$\arcmin\times$5$\arcmin$, in  Figure \ref{figure3}.
The age of the galaxy is 12.2\,Gyr. 
The stellar systems began to 
merge into a unique configuration about 8 \,Gyr ago.  The SFR proceeds 4.7\,Gyr before the galaxies merge and contributes to 24\% of the total SFR.
 The total  mass of gas  
inside a region of  50\,kpc  at the current galaxy age is 5\,$\times$ 10$^8\,M_\odot$ but no cold gas ($T\le$10000\,K) is expected (Figure \ref{figure2}).\\
\indent
\citet{Sietal10}  estimated a similar age (about 1--1.5\,Gyr old) of the 
stellar populations in the nuclear region  of this galaxy and  NGC~3626 \citep{paolanew}, both lenticulars.
The stellar nucleus in NGC~3599 at $R>$3$\arcsec$ is 5\,Gyr old at least.
We find that the average stellar age within $r_{e}$ is 4.5\,Gyr and  rises to 5.7\,Gyr within $R_{25} \simeq$ 2.4 $\times r_{e}$.
Ages weighted  by $B$-band luminosity become  younger, 3.4  and 4.8\,Gyr (Table  \ref{table3}), respectively, in good agreement with the previous findings by \citet{Sietal10}.
 At odds with us, \citet{Sietal10} suggested that this ETG results from a minor merger.  
Simulated and observed maps in Figure \ref{figure3} do not show any signature of interactions, in good agreement with the findings 
of \citet{Ant13}.\\
{\it{\object{NGC~3605.}}}~~~~~
A  merger  of  two triaxial systems with a mass ratio of 1:1 and  total mass  
in Table \ref{table2} matches  well with the $B$-band absolute magnitude,
the SED, and the morphology  of this dwarf E galaxy
(Figures \ref{figure1} and \ref{figure3}).
 The spins of the merging systems are equal ($\lambda$=0.06, MC03), 
perpendicular,  and both aligned with the shorter of their principal axes. 
The age of the galaxy  is 13.8 Gyr.
The SFR proceeds  about 3\,Gyr before the merger and contributes to 16\% of the total SFR.
 The average stellar age  within $r_{e}$ is 5\,Gyr  and becomes 6\,Gyr 
within $R_{25} \simeq$ 2$\times r_{e}$ (weighted by $B$-band luminosity: 3.4\,Gyr and 4.3\,Gyr, respectively).\\
\indent 
The system does not form stars at the selected snapshost, consistent with the 
overall picture of an old E  \citep{Noetal06, Ant13}.\\
\indent
The amount of cold  gas provided by the simulation is, on average, lower than  5$\times$10$^7\,M_\odot$ in the last stages of its evolution
 (Figure \ref{figure2}), in
agreement with \citet{Weletal10} who measured  less than 4.8$\times$10$^7\,M_\odot$. These authors also estimated  a  stellar mass of 
1.5$\times$10$^9\,M_\odot$
comparable with our value, 2$\times$10$^9\,M_\odot$ (from Table \ref{table3}).\\
{\it{\object{NGC~3607.}}}~~~~~
The simulated $B$-band magnitude (Table \ref{table3}), the SED (Figure \ref{figure1}), and the morphology 
(Figure \ref{figure3}) of  NGC~3607,   the  brightest member of U376
are well matched by the merging of  two triaxial 
collapsing  systems with a   mass ratio of 2:1 and a total mass almost 10 times larger than that of NGC~3605 (Table \ref{table2}). 
This is a direct encounter with a
first pericenter separation, $p$,  corresponding to 1/4 of the semi-major axis of 
the primary system. The stellar systems merged into a unique configuration about 8\,Gyr ago.
The current age of the galaxy is 13.8\,Gyr.  The SFR proceeds  about 6.5\,Gyr before the merger  and contributes to about
1/5 of the total SFR.
The average stellar age is younger than the galaxy age as a consequence
 of  the last  burst of SF occurring about 3.5\,Gyr ago. In  the inner ($r \le r_{e}$) region,  the average age of the stellar
component is almost  4\,Gyr and becomes slightly younger,  3\,Gyr, weighted by $B$-band luminosity. This value
agrees well with the luminosity-weighted age 
of 3.1$\pm$0.5\,Gyr measured by \citet{Annib10}  using Lick line-strength indices.  The same authors also estimated
a stellar metallicity of 0.047$\pm$0.012 (their Table 1), in good agreement with our average value,   0.044, inside
$r_{eff}$.\\
\indent
For this galaxy, the  observed SED extends up to FIR including new
AKARI/FIS data\footnote{Fluxes at 65 and 160 $\mu$m are upper limits} (Figure \ref{figure1}).
The predicted FIR SED is composed of a warm and a cold dust  
component both including polycyclic aromatic hyfrocarbon (PAH) molecules \citep{paola2, paola4}. 
Warm dust is located in regions of high radiation,  i.e., in the  neighborhood 
of OB clusters, whereas cold dust is heated by the  interstellar radiation field. 
The intensity and the distribution of diffuse radiation  field  in Figure \ref{figure1}
are the same as derived by \citet{paola4} for a nearby complete sample of ETGs. 
In particular, its warm dust is  cooler ($T \simeq$46~K instead of 61~K)
 and the warm-to-cold energy ratio is four times higher than the average value for ETGs in \citet{paola6}.\\
\indent
The galaxy has a large amount of molecular gas, log M(H$_2$)=9.13\,$M_\odot$
\citep{Temi09I}, which is  spatially extended \citep{Youngetal11}.
The simulation also indicates the presence of  a large amount of residual gas 
within a sphere  centered on the galaxy  (Figure \ref{figure2}).
The mass of  gas with a temperature $\le 2\times$10$^4$\,K, the
 upper limit of  cold gas mass (its cooling timescale is much 
shorter than the snapshot time range), is 1.5$\times$10$^9$ $M_\odot$ at the 
age of our fit, in agreement with   \citet{Temi09I}.\\
{\it{\object{NGC~3608.}}}~~~~~
The  merging of  two triaxial collapsing  systems with a mass ratio of 1:1 and a
 total mass  slightly larger than that of NGC~3607 (Table \ref{table2}) accounts for the global properties of this E.
Its simulated $M_B$ (Table \ref{table3}) agrees well with observations (Table \ref{table1}), as well as its SED 
and morphology (Figures \ref{figure1}  and \ref{figure3}).
The spins of the systems are equal 
($\lambda$=0.06) and both are aligned with the shorter of their principal axes.
They merged into a unique configuration about 12.5\,Gyr ago.  The SFR proceeded  about 1.3\,Gyr before the merger 
 and contributes to about 2\% of the total SFR.\\
\indent
The age of the galaxy  is 13.8\,Gyr. The average stellar age  within $r_{e}$ is about 5\,Gyr 
and becomes 5.5\,Gyr within 2$\times r_e$.
 Signatures of accretion rest on some photometric and kinematical peculiarities  \citep{McD06}. They estimate 
a galaxy age  in the inner region  of  8.9$^{+4.1}_{-2.8}$ Gyr.\\
\indent
The FIR SED takes into account the contributions of the warm and cold dust components, both including PAH molecules,  as described for NGC~3607.
The intensity and the distribution of diffuse radiation  field, which drives cold dust emission, are  as  in \citet{paola4}. 
The warm dust is  slightly cooler than  their average,  with a temperature of 
about 55~K  instead of  61~K. The warm-to-cold energy ratio is twice that of the 
average, i.e.,  1 instead of 0.5 as in \citet{paola6}.\\ 
\indent  
No residual  SFR  is predicted in NGC 3608, in agreement with the mid-infrared $Spitzer-Infrared-Spectrograph$ results
\citep{Rampazzo13}, which depict the galaxy as a passively evolving ETG.\\
\indent
The gas accretion history of this simulation (Figure 
\ref{figure2}) indicates  ~2.8$\times$10$^{8}\,M_\odot$ of gas within 50\,kpc.
 \citet{Weletal10} reported a cold gas  amount  of 1.4 $\times$10$^8\,M_\odot$,  in good agreement with our  upper limit
of  ~1$\times$10$^8\,M_\odot$ for the cold gas ($T \le$20000\,K).\\    
{\it{\object{UGC~06324.}}}~~~~~
The global properties of this  dwarf S0, at the outskirts of the U376 group (Figure \ref{dyn_a}),
are well matched by a  simulation of an encounter between two triaxial 
collapsing systems of equal mass and perpendicular spins.
Its SED, from FUV to the $K$ band, and morphology are shown in  Figures \ref{figure1} and \ref{figure3}, respectively.
\citet{Ant13} find no obvious signature of interaction  in  the  FUV,  NUV,  
or  in SDSS band images of this galaxy.\\
\indent
The  $B$-band absolute magnitude  (Table \ref{table3}) agrees well with 
the measured value in Table~\ref{table1}. The  total mass involved is 2$\times$10$^{11}\,M_\odot$ (Table \ref{table2}), but  that
of the galaxy is about 10 times less (Table \ref{table3}).
Its current  age is 10.8\,Gyr and the  average stellar age  within 
$R_{25}$  is 5\,Gyr ($\simeq$2\,Gyr when weighting stellar ages by $B$-band luminosity).\\
\indent
The gas accretion history  within 50\,kpc, corresponding  to $\simeq$12.5R$_{25}$, shows 
a large amount of gas,  ~1$\times$10$^{9}\,M_{\odot}$,  at  the current 
age  (the vertical line in Figure \ref{figure2}),
which reduces to 4$\times$10$^8\,M_{\odot}$ within $D_{25}$. No cold ($T\le$10000\,K) gas is expected, in agreement with  \citet{SengeBal06}.\\
{\it{\object{NGC~3626}}}~~~~~
This lenticular galaxy has been the subject of a previous paper \citep{paolanew}.
Its global properties are well matched by a simulation of a 2:1 merger with a total mass of 3 $\times$10$^{12}\,M_{\odot}$.

In summary, all the ETGs in U376 arise from a major merger except for the dwarf lenticular galaxy  UGC~06324 which
  is the result of a galaxy encounter.
Es have the same age,  older than 13\,Gyr, and  S0s  are younger than 1\,Gyr at least.
We note that, as pointed out in Section 3,  the total mass of all simulated systems in  U376   does not exceed the virial mass of the group  
(Table  \ref{tableA1})  from the beginning (Table \ref{table2}), fulfilling the condition set in Section 2.

\subsection{LGG~225: Individual Notes of its ETGs}

\begin{figure*}
\centering
{\includegraphics[width=10cm]{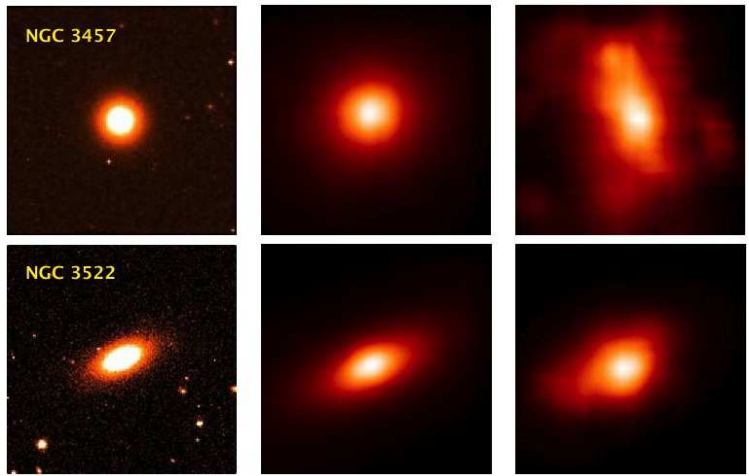}}
     \caption{As in Figure \ref{figure3} for ETGs in  LGG~225.
}
   \label{figure4}
 \end{figure*}

\medskip\noindent
\underbar{\object{NGC~3457}}~~~~~
This E is well matched  by a  
simulation of an encounter between  two triaxial collapsing halos 
of equal mass and perpendicular spins, like UGC~06324, but with a total mass
10 times larger than that of  UGC~06324 (Table \ref{table2}). 
 The  predicted $B$-band absolute magnitude,  $M_B=-18.54$\,mag (Table \ref{table3}), 
agrees well with the measured value in Table \ref{table1}.
 All the total photometric data, from FUV to 100\,$\mu$m, are well matched by the
SED provided by our simulation at the selected snapshot (Figure \ref{figure1}).
Warm and cold dust components in the FIR SED, where only upper limits are available,  have the same properties as on
average for Es in \citet{paola4} and  \citet{paola6}.
The current age of the simulated galaxy is 11.4\,Gyr and its average stellar age within R$_{25}$,  weighted by $B$-band luminosity, is 4\,Gyr
 (Table  \ref{table3}).\\
\indent
\citet{Ant10} derived a stellar mass of 2--4$\times$10$^9$\,$M_{\odot}$ from their SED fit
extending from FUV to $K$ band.  This value is  within a factor of two from our value, 
9.6$\times 10^{9}$\,$M_{\odot}$, although based on both  different lower mass limit of the IMF and SFR \citep{Antonew}.\\
\indent
Figure \ref{figure4} (top panel, left and middle columns) compares, on the same scale and with the same
 resolution, 
the  SDSS $r$-band image of  this galaxy with that of our best-fit snapshot.\\
\indent
The gas accretion history within a sphere of radius 50\,kpc  ($\sim$16.4$\times R_{25}$) indicates 
 1.5$\times$10$^{8}$\,$M_{\odot}$ of gas at the current age. 
 \citet{Danietal03}  report 10$^8$\,$M_\odot$ of HI in this galaxy, in good agreement with our predictions.\\ 
\indent
Accounting for the further constraint put in Section 2, based on the dynamical analysis of this group \citep[][see Appendix]{Antonew},
this simulation provides a total mass of
5.7$\times$10$^{11}$\,$M_{\odot}$   within a radius corresponding to its
virial radius (1.5$\times R_H$ in Table \ref{tableA1}), at the current age.\\
\medskip\noindent
\underbar{\object{NGC~3522}}~~~~~
A merging of two halos with a mass ratio of 1:1  and a total mass five times less than that of NGC~3457
(Table \ref{table2})  accounts for the evolution of this
dwarf ETG.  The spins of the systems are equal ($\lambda$=0.06; MC03), 
parallel, and both are aligned  with the shorter of their principal axes. 
They merge into a unique configuration about 9.4\,Gyr ago, the current age of the simulated galaxy 
being 12.6 Gyr.
 The SFR proceeds  about 3.1\,Gyr before the merger 
 and contributes to about 16\% of the total SFR.
The $B$-band total magnitude of the simulated galaxy is -17.81\,mag,
 in good agreement with the values in Table \ref{table1}. The SED provided by our simulation
best fits all the available total fluxes (Figure \ref{figure1}).
Warm and cold dust components in the FIR SED have the same properties as on average for Es  in \citet{paola4} and \citet{paola6}.
In particular, the fraction of the bolometric luminosity in the FIR range is 0.017, and the luminosity fraction of warm dust 
is 0.3$\%$ of the FIR luminosity.
The average stellar age within R$_{25}$ is  4.7\,Gyr   and
 3.3\,Gyr weighted  by $B$-band luminosity (Table \ref{table3}).\\
\indent
 \citet{Ant10} estimated a total stellar mass of 1--2$\times$10$^9\,M_{\odot}$  from their SED fit 
extending from FUV to $K$ band. 
Their value agrees well with our value, 3.4$\times\,10^{9}\,M_{\odot}$,  although they used a different 
lower mass limit for the IMF  and a different SFR \citep{Antonew}.\\
\indent
Figure \ref{figure4} (bottom panel, left and middle columns) compares the
 SDSS $r$-band image with that of our best-fit snapshot. Figure \ref{figure2} 
 shows the  gas  mass evolution within a radius of 50\,kpc (i.e., $\simeq$13 $\times\,R_{25}$) from 
the $B$-band  galaxy center.
There is a large amount of gas, 10$^9\,M_{\odot}$, inside such a radius at the selected snapshot, 
 but the mass of cold gas within 2$\times\,R_{25}$ is at least 10 times 
less,  in good agreement with \citet{Weletal10} which report a value of 1.1$\times$10$^8$\,M$_{\odot}$. 
They classified this galaxy as  a gas-rich ETG.\\
\indent
The total mass  within  the virial radius of the group (1.5$\times$R$_H$ in Table \ref{tableA1}) at the current  age 
is  2.7$\times\,10^{11}\,M_{\odot}$.\\

\indent
Es in LGG~225 are younger than those in U376 of 1\,Gyr at least. NGC~3457 is the result of a galaxy encounter and NGC~3522
arises from a major merger, whereas all the Es in U376 derive from a major merger.\\
Accounting for previous results,
the total mass of the simulations performed to fit the global
properties of ETGs within the viral radius  of the LGG~225 group (Table  \ref{tableA1}), is 
8.4$\times\,10^{11}\,M_{\odot}$. Therefore, both simulations fulfill the condition set in Section 2.\\

\section{Following the galaxy transformation. \\
The (NUV-$r$) vs. $M_r$  CMD diagram of U376 and LGG~225}

\begin{figure*}
\centering
\includegraphics[width=13cm]{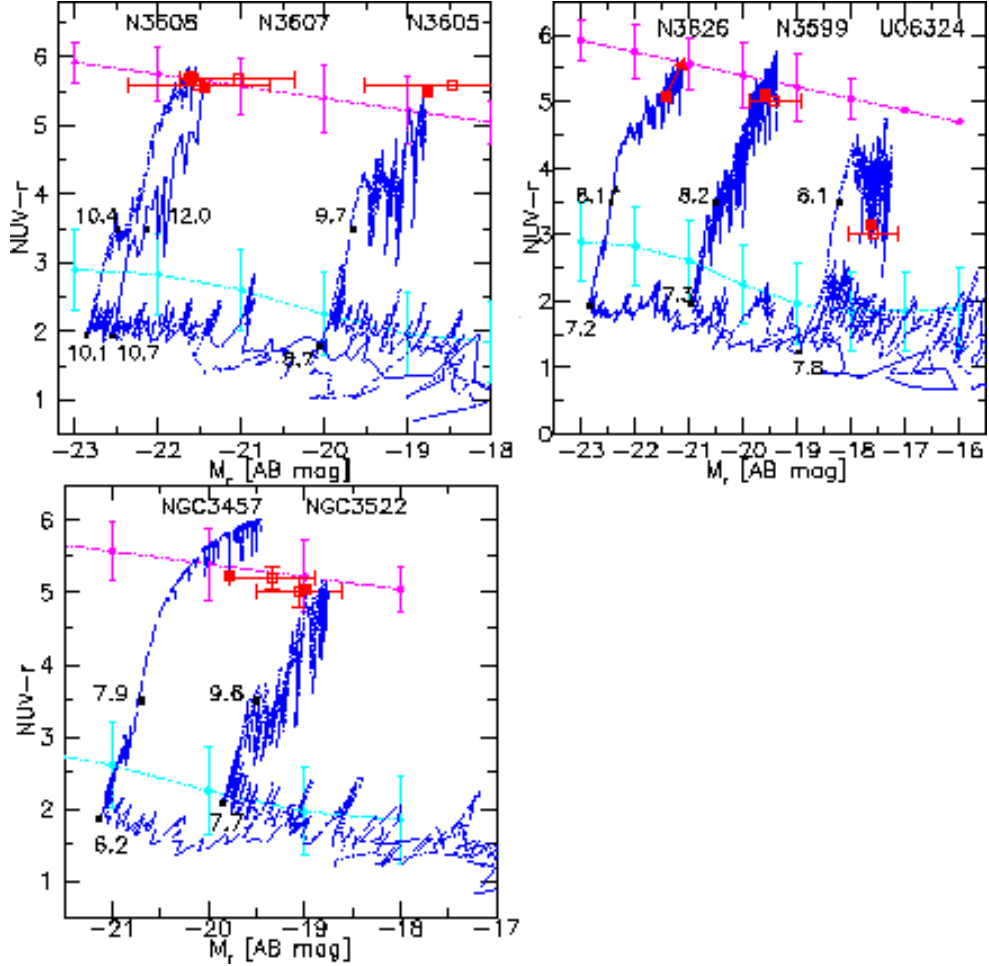}
 \caption{Rest-frame  NUV-$r$ vs. $M_r$ CMD of ETGs separated in Es (left) and S0 (right) for U376 (top) and LGG~225 (bottom).
  Red filled symbols  correspond to the current age of ETGs from our simulations in Table \ref{table3}.
 The measured values are shown with empty symbols accounting for the average distance moduli between NED (3K CMB) values and observed ones, when available, and
the corresponding 3\,$\sigma$ uncertainties.  
The RS (magenta) and the BC (cyan) are  also plotted following prescriptions in \citet{Wyderetal07}.}
\label{figure5}
\end{figure*}

NUV-$r$ color is an excellent tracer of even small amounts 
($\simeq$1\% mass fraction) of recent ($\le$1 Gyr) SF \citep[e.g.][see also Section 1]{Salim05, Schaw07}.
By studying  a large sample of local (redshift from 0 to 0.11) ETGs,  \citet{Kavi07} concluded that at least 30\% 
 of them have NUV-$r$  colors consistent with some recent  ($\le$1\,Gyr) SF.\\
Our fully consistent chemo-photometric simulations permit us to follow not only the NUV-$r$  color 
variation in the  CMD but also the galaxy morphological transformation. 
 Observations of clusters  at 0.2 $\leq z \leq $ 0.7 \citep{Fasano2000} show that a sort of conversion in galaxy
population from S to  S0s takes place at  $z \approx$ 0.3--0.4, corresponding to 
3--5\,Gyr ago. The mechanisms driving the S-to-S0 transformation are still uncertain and debated \citep[e.g.][]{Bekki09}.
Recently, \citet{paolanew} analyzed the  galaxies NGC~1533 and NGC~3626, two
Sa/S0 galaxies showing ring/arm-like structures 
detected in H$\alpha$ and FUV. The two galaxies are located in nearby groups
and, in particular NGC~3626 is a member of U376.
They show that these two galaxies have a similar evolution, driven by
a major merging. The star-forming ring/arm-like structures
observed in the above galaxies  arise in the latter stages of the merger episode, due to a head-on encounter,
when the galaxies are almost 8\,Gyr old. \\
Figure \ref{figure5} shows the evolutionary paths of ETGs in our groups in the rest-frame 
CMD, resulting from theSPH simulations described  above. 
The evolution of ETGs is stopped  at 14\,Gyr. Some significant 
evolutionary stages are emphasized with dots and bigger dots  correspond to their current age.
We assume 3.5 as the NUV-$r$ threshold of the GV,  this value marks the GV entry of 
the brightest galaxy in our sample. We assume   5 for the RS 
threshold.
Several ``rejuvenation" episodes, which  appear in the 
CMD  as oscillations (see Figure \ref{figure5}), make or keep the color bluer than the RS threshold. 
For example, NGC~3457 is about 1\,Gyr younger than NGC~3522 (Table \ref{table3});
however its stellar populations are, in average, 1\,Gyr older. NGC~3522 gets in the GV 
after $\approx$10\,Gyr instead of 8\,Gyr as in NGC~3457. 
Then, both the galaxies reach their current position on the RS in about 3\,Gyr. 
The path in the GV  of NGC~3522 is marked by several rejuvenation
episodes, at odds with that of NGC~3457.
Therefore, the  optical luminosity of NGC~3457 at the current age is dominated by stellar populations  older  than those in NGC~3522. 
The rest-frame CMD of NGC~3607 and  NGC~3605 shows rejuvenation episodes  that occurred while crossing 
the GV, 1\,Gyr ago.
In all the S0 galaxies in our sample several rejuvenation episodes  
occurred in the last 1\,Gyr  (Figure \ref{figure5}).\\
\indent
The  morphological transformation of selected galaxies in U376 and LGG~225 happened 
3.9, 5.1, 3.1, 3.7, and 3.7\,Gyr ago for NGC~3599, NGC~3605, NGC~3607, NGC~3608, and UGC~06324, respectively, 
and 5.2 and 4.8\,Gyr for NGC~3457 and NGC~3522, respectively. We show the morphology at these lookback times, 
(corresponding to their brightest absolute magnitude in the CMD of Figure \ref{figure5}),
in Figure \ref{figure3} and  \ref{figure4} (right panels) for U376 and LGG~225 members respectively. \\
\indent
From the previous analysis,  we conclude that:
(1) Es with  current magnitudes brighter than $M_r$[AB]= -21\,mag, lived the 
BC more quickly the brighter they are. They reach their current 
position on the RS, 2--4\,Gyr later, showing few and/or weak fluctuations both in the GV and/or in the RS. 
(2) Es  whose current magnitudes are lower than $M_r$[AB]$= -21$\,mag, 
left the BC 2--4\,Gyr before their brighter counterparts, show more oscillations in the GV, and
reach their current  position on the RS  almost 5\,Gyr later. 
(3) S0 galaxies in U376 moved  through the  BC more quickly than Es and they moved more quickly the brighter they are. The brighter S0s
show  fewer and weaker fluctuations in the GV.
S0s may either cross the GV and reach the RS in the same time ($\approx$2--4\,Gyr) as well as Es, like
 NGC~3626, or  spend all their remaining lifetime oscillating from the GV to the RS or vice versa, as in the case of UGC6324, depending on their SF  histories.\\
Therefore, fainter ETGs  may experience rejuvenation episodes more frequently than other ETGs.

\section{Summary and Conclusions}

 We  explore  the merger/interaction scenario  starting  from collapsing triaxial systems composed of DM and gas
and combine the SPH code with the chemo-photometric code based on EPS models.
 We trace back the evolution of  ETGs in two groups  of the Leo  cloud, U376 and LGG~225, using  
 chemo-photometric SPH simulations that are able to match their current global properties.
{\bf  Simulations of isolated collapsing  halos, in the same framework  of parameters, indeed, do not reach the same goal.}\\
\indent
Here we focus on their SED, morphology, and $B$-band absolute magnitude, and on
their evolutionary path  in the rest-frame, NUV--$r$ $versus$ $M_r$ CMD.
This diagram is a powerful diagnostic tool to detect very low levels of SF.
The transition from the BC, where the analyzed ETGs  stay at least 7\,Gyr, across the GV, 
lasts about 3--5\,Gyr. 
 \citet{Fasano2000} found that a sort of conversion in 
galaxy population from S to  S0s takes place at  $z \approx$ 0.3--0.4, corresponding to 3--5\,Gyr ago,
in good agreement with our findings \citep[see also][]{paolanew} .\\
\indent
Excursions into the GV from the RS, driven by the acquisition of fresh
gas for SF,   are more numerous for dwarf ETGs, so
fainter ETGs may experience rejuvenation episodes more frequently than the brightest, more massive ones.
Some Es in our sample, namely NGC~3607 and NGC~3605 in U367, and 
NGC~3522 in LGG~225, experienced rejuvenation episodes within the last 1\,Gyr 
during their walk through the GV, in agreement with the findings of  \citet{Kavi07}.\\
\indent
The evolutionary history of most of  the  ETGs are well matched 
by a {\it single wet major merger episode} of two halos. SF episodes are regulated by
the secondary episodes of gas infall.
The S0 galaxy U06324 in U376, and the E galaxy NGC~3457 in LGG~225 have the SF  quenched by a close encounter 
with a companion. \\
Within the above framework, we suggest that at least one of the 
mechanisms of  the morphological and photometric  transformation of S
 into S0s in groups, is gravitational, unrelated to the 
intragroup hot  medium \citep{paolanew}.\\
\indent
We find that the ETGs in U376 and LGG~225  have followed 
different evolutionary paths and experienced merger phenomena. The ETGs have come 
  from different areas of the cosmic web, within the virial radius of this 
group. This picture agrees well with the findings of  \citet{Wetzeletal12}.  
They found   no significant environmental effects on galaxies before they cross  the $R_{vir}$ 
 of a more massive host halo, which, in our case, would be built up by the 
 same galaxies we see now, at their current age.\\
\indent
Es in LGG~225 are younger than those in U376 by 1\,Gyr at least. NGC~3457 is the result of a galaxy encounter and NGC~3522
arises from a major merger, whereas all the Es in U376 derive from a major merger.\\
The results of our simulations agree  well with the dynamical analysis of these groups,
which puts LGG~225 in a pre-virialized phase and U376 in a more evolved phase, toward virialization although not  virialized yet \citep{Antonew}.

\acknowledgments

A.M. and R.R. acknowledge the financial contribution from the agreement ASI-INAF I/009/10/0. 
This work has been partially supported by the Padova University funds 2011/12 (ex 60\%).
This research has made use  of the HyperLeda (http://leda.univ-lyon1.fr) \citep{Paturel03}  and the NASA/IPAC Extragalactic Database (NED),
which is operated by the Jet Propulsion Laboratory, California Institute of Technology, under contract with the National Aeronautics and Space 
Administration.\\ 
{\it Facilities: {\it  GALEX}, and SDSS}
\clearpage
\appendix

\section{Summary of the  Dynamical Analysis of U376 and LGG 225}
\begin{figure*}
  \centering
 {\includegraphics[width=17cm]{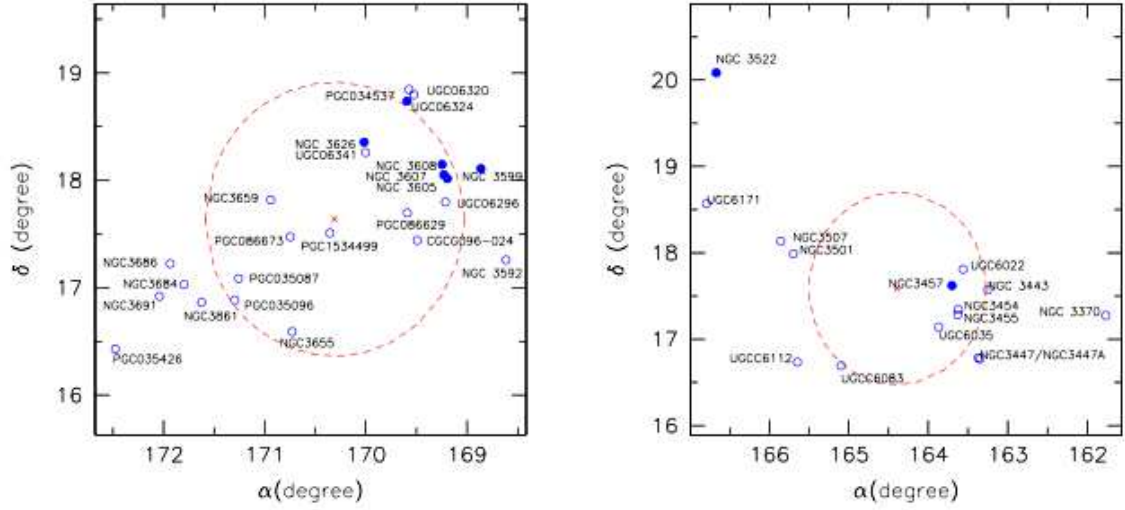}}  
\caption{ 
     Projected distribution of the positions of members in U376 (left) \citep{Ant13} and LGG~225 (right).
    Filled  (blue) circles  are ETGs.
    The virial radius (dashed line; see Table \ref{tableA1}) of each group is also shown.
  ETGs are well outside the central region in both the groups, a clue that groups are not virialized yet.}
     \label{dyn_a}
   \end{figure*}

Summary of the Kinematical and Dynamical Analysis of U376 and LGG 225\\
\indent
We performed a kinematical and dynamical analysis of U376 and LGG~225, whose membership  
haa been fully defined in previous papers \citep{Ant10, Ant13},  accounting for the recipes 
given by \citet{Peretal90} and \citet{Firth06}. As discussed in these papers, a large scatter in the 
properties of the group could arise
from this  approach depending on if we weight  by luminosity or not, and on the adopted 
 wavelength \citep{Ant10, Bettoni11}, due
to the different distributions of mass and light.\\
\indent
\citet{Antonew} presented the  non-luminosity-weighted dynamical properties of U376  using 
$H_{0}$=75\,km\,s$^{-1}$\,Mpc$^{-1}$.
Here, we apply  the same recipes with $H_{0}$=70\,km\,s$^{-1}$\,Mpc$^{-1}$ (Section 1). 
Errors are derived via the jackknife method \citep{Efron82}.  Table \ref{tableA1} summarizes our 
results which are the same, within the errors, of the previous paper, showing that the H$_{0}$ value
 adopted here does not produce a meaningful change in the  dynamical results.
Figure \ref{dyn_a} emphasizes the positions of our selected ETGs. 
In both the groups, the ETGs appear far away from the dynamical center of the group. Moreover,
accounting for  the presence of substructures  in U376 \citep{Ant13}
and the large crossing-time  characterizing LGG~225 (Table \ref{tableA1}), probably both groups are not  virialized yet.

 \clearpage


\clearpage
\begin{table*}
\scriptsize
\caption{Properties of ETGs in U376 and LGG~225}
\begin{tabular}{llcccccc}
\hline\hline
Group &Galaxy & Type  &scale$^a$  &D$_{25}$ &r$_{e}$ &B& M$_B$\\
          &ETG &  &(kpc\,\arcsec$^{-1}$  &(kpc)&(kpc)&(mag)&  Range (mag)\\
\hline
& & & & & & & \\
{ U376} &   & & &  & & &\\
& & & & & & &\\ 
\hline
& NGC~3599  &S0   &0.081  &11.6 & 2.43$^1$&12.78$\pm$0.08&-17.91   -18.33  \\
& NGC~3605  &E    &0.072  & 5.4 & 1.24$^1$--1.51&13.06$\pm$0.28&-17.23   -17.81 \\
& NGC~3607  &E-S0 &0.089  &24.4 & 3.9$^2$--5.8$^1$ &10.83$\pm$0.17&-20.04   -20.51\\
& NGC~3608  &E    &0.110  &20.2 & 3.7$^1$--4.3$^3$&11.46$\pm$0.26& -19.93   -20.32 \\
& NGC~3626  &S0-a &0.127  &22.3 & 3.3 &11.70$\pm$0.23&-20.11 -20.38  \\
& UGC~06324 &S0   &0.097  &8.0  & $--$    &14.66$\pm$0.32   &-16.52   -17.20\\
\hline
& & & & & & & \\
 { LGG~225} &  & & &  & & &\\
& & & & & & &\\ 
\hline
& NGC~3457&E &0.104 &6.1 & --&12.8$\pm$0.24 &-18.55   -18.84\\
& NGC~3522&E &0.107  &7.5 & --&13.9$\pm$0.28& -17.49   -17.84 \\
\hline
\end{tabular}
\\
{\bf References}. {(1)~\citet{Fabetal89}, (2)~\citet{Marino11a}, (3)~\citet{SAURONIV}}\\

\label{table1}
\end{table*}

\begin{table*}
\scriptsize
\caption{Parameters of SPH Simulations of ETGs in U376 and LGG~225}
\begin{tabular}{llcccccccccc}
\hline\hline
Group & Galaxy & N$_{part}$(t=0)&a & $p/a$ & $r_1$ & $r_2$ & $v_1$ & $v_2$ & $M_T$\\
&ETG &  &(kpc)&   &(kpc) & (kpc)& (km/s) & (km/s)  & ($10^{10}\,M_\odot$)  \\
\hline
& & & & & & & & &   \\
{ U376} & & & & & & &  & &  \\
& & & & & & & & & & \\
\hline
& NGC~3599 &8$\times$10$^4$&597& 1/4  & 343  & 343 &38&38& 80   \\
& NGC~3605 &8$\times$10$^4$&474& 1/5  & 218  & 218 &34&34& 40 \\
& NGC~3607 &6$\times$10$^4$&1014 & 1/4  & 390  & 778 &38 & 76 & 300  \\
& NGC~3608 &6$\times$10$^4$&1014& 1/10  & 263  & 263 &104 &104& 400 \\
& NGC~3626 &6$\times$10$^4$&1014& 1/10 & 273  & 546 &52 & 104 & 300   \\
& UGC06324 &8$\times$10$^4$&376 &1/3  & 506  & 506 & 18 &18& 20 \\
\hline
& & & & & & & & & & & \\
{ LGG~225} & & & & & & & & & & & \\
& & & & & & & & & & & \\
\hline
& NGC~3457 &8$\times$10$^4$&810 & 1/3 &621&621 & 45  & 45  & 200   \\
& NGC~3522 &8$\times$10$^4$&474&  1/5&218 &218  & 34.2 & 34.2  & 40  \\
\hline
\end{tabular}
\label{table2}
\end{table*}

\begin{table*}
\scriptsize
\caption{Results within $R_{25}$ of ETGs in U376 and LGG~225}
\begin{tabular}{llcccccccc}
\hline\hline
Group & Galaxy& $t_{gal}$ &$t_{pop}$&$Z_{*}$  & $M_{tot}$& $M/L_B$ & $f_{DM}$ &  $f_{gas}$  & $M_B$\\
&ETG  &(Gyr)& (Gyr) & & (10$^{10}\,M_{\odot}$)& ($M_{\odot}$/$L_\odot$) & & &(mag)\\
\hline
& & & & & & & & & \\
{ U376} & & & &  & & & & & \\
& & & & & & &  & & \\
\hline
& NGC~3599&12.2   & 4.8& 0.013 & 1.8 & 22.7 & 0.41  &0.01 &-18.37 \\
& NGC~3605&13.8   & 3.8& 0.012 &0.29 &  14.5& 0.23 & 0.05& -17.57 \\
&NGC~3607&13.8  & 4.0 & 0.015&  11.2&17.6 & 0.30 & 0.01 &-20.25 \\
&NGC~3608&13.8   & 6.0 &0.018 &  10.9& 16.9 & 0.31 & 0.002&-20.37\\
&NGC~3626&11.5   & 4.5 & 0.016&11.5& 18.8 & 0.19  &0.08 & -20.23 \\
& UGC~6324&10.8   & 1.9 & 0.005 &0.16& 6.3 & 0.24 &0.24& -16.75 \\
\hline
& & & & & & & & \\
{ LGG~225} & & & & & & & & \\
& & & & & & & & \\
&NGC~3457&11.4  & 4.3&  0.010& 0.15& 18.2 & 0.35 & 0.03 &-18.54 \\
& NGC~3522&12.5   &3.3&0.019   &0.48 & 11.7 & 0.26 & 0.03&-17.80\\
\hline
\end{tabular}
\label{table3}
\end{table*}

\begin{table*}
\scriptsize
\caption{Dynamical Properties of U376 and LGG~225}
 \begin{tabular}{llcccccccc}
\hline\hline
Group & Center &  $V_{group}$  & Velocity    & $D$   & Harmonic &Virial  & Projected      & Crossing \\
 Name & of Mass  &             & Dispersion &    &    Radius      & Mass   &  Mass   & Time$\times\,H_0$  \\\
     & RA [deg]  ~ Dec (deg) &(km/s) &(km/s)& (Mpc)  & (Mpc)     &  (10$^{13}\,M_{\odot}$) &  (10$^{13}\,M_{\odot}$)  & \\ 
\hline
U376       & 170.3058  17.6398  & 1134$^{+14}_{-19 }$  &   225$^{+5 }_{-19 }$ &   16.20$^{+0.27 }_{-0.30 }$ &0.23$^{+0.02}_{-0.01}$     &  1.28$^{+0.18}_{-0.22}$ &  3.12$^{+0.24}_{-0.45}$ &  0.13$^{+0.02}_{-0.01}$\\

LGG~225    & 164.3925  17.5876  & 1100$^{+12}_{-13}$    &  92.2$^{+3.2}_{-10.9}$   & 15.72$^{+0.17}_{-0.18}$&  0.20$^{+0.05}_{-0.03}$ &   0.18$^{+0.07}_{-0.01}$ &  1.02$^{+0.09}_{-0.31}$ & 0.37$^{+0.05}_{-0.04}$ \\
\hline
\end{tabular}
\label{tableA1}
\end{table*}
\clearpage
 \bibliographystyle{apj}
 \bibliography{U376_NUVr_evol-rev3}

\begin{thebibliography}{77}
\expandafter\ifx\csname natexlab\endcsname\relax\def\natexlab#1{#1}\fi

\bibitem[{{Annibali} {et~al.}(2010){Annibali}, {Bressan}, {Rampazzo},
  {Zeilinger}, {Vega}, \& {Panuzzo}}]{Annib10}
{Annibali}, F., {Bressan}, A., {Rampazzo}, R., {et~al.} 2010, \aap, 519, A40

\bibitem[{{Baldry} {et~al.}(2004){Baldry}, {Glazebrook}, {Brinkmann},
  {Ivezi{\'c}}, {Lupton}, {Nichol}, \& {Szalay}}]{Baldry04}
{Baldry}, I.~K., {Glazebrook}, K., {Brinkmann}, J., {et~al.} 2004, \apj, 600,
  681

\bibitem[{{Ball} {et~al.}(2008){Ball}, {Loveday}, \& {Brunner}}]{Ball08}
{Ball}, N.~M., {Loveday}, J., \& {Brunner}, R.~J. 2008, \mnras, 383, 907

\bibitem[{{Balogh} {et~al.}(2004){Balogh}, {Baldry}, {Nichol}, {Miller},
  {Bower}, \& {Glazebrook}}]{Balogh04}
{Balogh}, M.~L., {Baldry}, I.~K., {Nichol}, R., {et~al.} 2004, \apjl, 615, L101

\bibitem[{{Bekki}(2009)}]{Bekki09}
{Bekki}, K. 2009, \mnras, 399, 2221

\bibitem[{{Bekki} \& {Couch}(2011)}]{Bekki11}
{Bekki}, K., \& {Couch}, W.~J. 2011, \mnras, 415, 1783

\bibitem[{{Bettoni} {et~al.}(2012){Bettoni}, {Buson}, {Mazzei}, \&
  {Galletta}}]{Dani12}
{Bettoni}, D., {Buson}, L., {Mazzei}, P., \& {Galletta}, G. 2012, \mnras, 423,
  2957

\bibitem[{{Bettoni} {et~al.}(2003){Bettoni}, {Galletta}, \&
  {Garc{\'{\i}}a-Burillo}}]{Danietal03}
{Bettoni}, D., {Galletta}, G., \& {Garc{\'{\i}}a-Burillo}, S. 2003, \aap, 405,
  5

\bibitem[{{Bettoni} {et~al.}(2011){Bettoni}, {Galletta}, {Rampazzo}, {Marino},
  {Mazzei}, \& {Buson}}]{Bettoni11}
{Bettoni}, D., {Galletta}, G., {Rampazzo}, R., {et~al.} 2011, \aap, 534, A24

\bibitem[{{Binggeli} \& {Hascher}(2007)}]{BH07}
{Binggeli}, B., \& {Hascher}, T. 2007, \pasp, 119, 592

\bibitem[{{Blanton} \& {Berlind}(2007)}]{BlaeBe07}
{Blanton}, M.~R., \& {Berlind}, A.~A. 2007, \apj, 664, 791

\bibitem[{{Blanton} {et~al.}(2005){Blanton}, {Eisenstein}, {Hogg}, {Schlegel},
  \& {Brinkmann}}]{Bla05}
{Blanton}, M.~R., {Eisenstein}, D., {Hogg}, D.~W., {Schlegel}, D.~J., \&
  {Brinkmann}, J. 2005, \apj, 629, 143

\bibitem[{{Cappellari} {et~al.}(2006){Cappellari}, {Bacon}, {Bureau}, {Damen},
  {Davies}, {de Zeeuw}, {Emsellem}, {Falc{\'o}n-Barroso}, {Krajnovi{\'c}},
  {Kuntschner}, {McDermid}, {Peletier}, {Sarzi}, {van den Bosch}, \& {van de
  Ven}}]{SAURONIV}
{Cappellari}, M., {Bacon}, R., {Bureau}, M., {et~al.} 2006, \mnras, 366, 1126

\bibitem[{{Christensen} {et~al.}(2012){Christensen}, {Quinn}, {Governato},
  {Stilp}, {Shen}, \& {Wadsley}}]{Chrietal2012}
{Christensen}, C., {Quinn}, T., {Governato}, F., {et~al.} 2012, \mnras, 425,
  3058

\bibitem[{{Christensen} {et~al.}(2010){Christensen}, {Quinn}, {Stinson},
  {Bellovary}, \& {Wadsley}}]{Chrietal2010}
{Christensen}, C.~R., {Quinn}, T., {Stinson}, G., {Bellovary}, J., \&
  {Wadsley}, J. 2010, \apj, 717, 121

\bibitem[{{Curir} \& {Mazzei}(1999)}]{CM99}
{Curir}, A., \& {Mazzei}, P. 1999, \na, 4, 1

\bibitem[{{de Vaucouleurs} {et~al.}(1991){de Vaucouleurs}, {de Vaucouleurs},
  {Corwin}, {Buta}, {Paturel}, \& {Fouqu{\'e}}}]{deV91}
{de Vaucouleurs}, G., {de Vaucouleurs}, A., {Corwin}, Jr., H.~G., {et~al.}
  1991, {Third Reference Catalogue of Bright Galaxies. Volume I: Explanations
  and references. Volume II: Data for galaxies between 0$^{h}$ and 12$^{h}$.
  Volume III: Data for galaxies between 12$^{h}$ and 24$^{h}$.}

\bibitem[{{Efron}(1982)}]{Efron82}
{Efron}, B. 1982, {The Jackknife, the Bootstrap and other resampling plans}

\bibitem[{{Faber} {et~al.}(1989){Faber}, {Wegner}, {Burstein}, {Davies},
  {Dressler}, {Lynden-Bell}, \& {Terlevich}}]{Fabetal89}
{Faber}, S.~M., {Wegner}, G., {Burstein}, D., {et~al.} 1989, \apjs, 69, 763

\bibitem[{{Fang} {et~al.}(2012){Fang}, {Faber}, {Salim}, {Graves}, \&
  {Rich}}]{Fangetal12}
{Fang}, J.~J., {Faber}, S.~M., {Salim}, S., {Graves}, G.~J., \& {Rich}, R.~M.
  2012, \apj, 761, 23

\bibitem[{{Fasano} {et~al.}(2000){Fasano}, {Poggianti}, {Couch}, {Bettoni},
  {Kj{\ae}rgaard}, \& {Moles}}]{Fasano2000}
{Fasano}, G., {Poggianti}, B.~M., {Couch}, W.~J., {et~al.} 2000, \apj, 542, 673

\bibitem[{{Firth} {et~al.}(2006){Firth}, {Evstigneeva}, {Jones}, {Drinkwater},
  {Phillipps}, \& {Gregg}}]{Firth06}
{Firth}, P., {Evstigneeva}, E.~A., {Jones}, J.~B., {et~al.} 2006, \mnras, 372,
  1856

\bibitem[{{Garcia}(1993)}]{Garcia93}
{Garcia}, A.~M. 1993, \aaps, 100, 47

\bibitem[{{Gonzalez} {et~al.}(2013){Gonzalez}, {Sivanandam}, {Zabludoff}, \&
  {Zaritsky}}]{Gonzetal2013}
{Gonzalez}, A.~H., {Sivanandam}, S., {Zabludoff}, A.~I., \& {Zaritsky}, D.
  2013, \apj, 778, 14

\bibitem[{{Gonz{\'a}lez-Samaniego} {et~al.}(2013){Gonz{\'a}lez-Samaniego},
  {Col{\'{\i}}n}, {Avila-Reese}, {Rodr{\'{\i}}guez-Puebla}, \&
  {Valenzuela}}]{GonSam13}
{Gonz{\'a}lez-Samaniego}, A., {Col{\'{\i}}n}, P., {Avila-Reese}, V.,
  {Rodr{\'{\i}}guez-Puebla}, A., \& {Valenzuela}, O. 2013, arXiv:
  astro-ph/1308.4753

\bibitem[{{Hogg} {et~al.}(2004){Hogg}, {Blanton}, {Brinchmann}, {Eisenstein},
  {Schlegel}, {Gunn}, {McKay}, {Rix}, {Bahcall}, {Brinkmann}, \&
  {Meiksin}}]{Hogg04}
{Hogg}, D.~W., {Blanton}, M.~R., {Brinchmann}, J., {et~al.} 2004, \apjl, 601,
  L29

\bibitem[{{Just} {et~al.}(2010){Just}, {Zaritsky}, {Sand}, {Desai}, \&
  {Rudnick}}]{Just10}
{Just}, D.~W., {Zaritsky}, D., {Sand}, D.~J., {Desai}, V., \& {Rudnick}, G.
  2010, \apj, 711, 192

\bibitem[{{Kauffmann} {et~al.}(2004){Kauffmann}, {White}, {Heckman},
  {M{\'e}nard}, {Brinchmann}, {Charlot}, {Tremonti}, \& {Brinkmann}}]{Kauff04}
{Kauffmann}, G., {White}, S.~D.~M., {Heckman}, T.~M., {et~al.} 2004, \mnras,
  353, 713

\bibitem[{{Kaviraj} {et~al.}(2007){Kaviraj}, {Schawinski}, {Devriendt},
  {Ferreras}, {Khochfar}, {Yoon}, {Yi}, {Deharveng}, {Boselli}, {Barlow},
  {Conrow}, {Forster}, {Friedman}, {Martin}, {Morrissey}, {Neff},
  {Schiminovich}, {Seibert}, {Small}, {Wyder}, {Bianchi}, {Donas}, {Heckman},
  {Lee}, {Madore}, {Milliard}, {Rich}, \& {Szalay}}]{Kavi07}
{Kaviraj}, S., {Schawinski}, K., {Devriendt}, J.~E.~G., {et~al.} 2007, \apjs,
  173, 619

\bibitem[{{Kawata} \& {Mulchaey}(2008)}]{Kawata08}
{Kawata}, D., \& {Mulchaey}, J.~S. 2008, \apjl, 672, L103

\bibitem[{{Kroupa}(2012)}]{K12}
{Kroupa}, P. 2012, \pasa, 29, 395

\bibitem[{{Krumholz} \& {Tan}(2007)}]{KrueTan07}
{Krumholz}, M.~R., \& {Tan}, J.~C. 2007, \apj, 654, 304

\bibitem[{{Lewis} {et~al.}(2002){Lewis}, {Balogh}, {De Propris}, {Couch}, \&
  et~al.}]{Lewis02}
{Lewis}, I., {Balogh}, M., {De Propris}, R., {Couch}, W., \& et~al. 2002,
  \mnras, 334, 673

\bibitem[{{Marino} {et~al.}(2014){Marino}, {Bianchi}, {Mazzei}, \& {et
  al}.}]{Antonew}
{Marino}, A., {Bianchi}, L., {Mazzei}, P., \& {et al}. 2014,
  arXiv:astro-ph/1309.5031, Advanced in Space Research, Special Iusses:
  Ultraviolet Astrophysics, accepted, 53

\bibitem[{{Marino} {et~al.}(2010){Marino}, {Bianchi}, {Rampazzo}, {Buson}, \&
  {Bettoni}}]{Ant10}
{Marino}, A., {Bianchi}, L., {Rampazzo}, R., {Buson}, L.~M., \& {Bettoni}, D.
  2010, \aap, 511, A29

\bibitem[{{Marino} {et~al.}(2011{\natexlab{a}}){Marino}, {Bianchi}, {Rampazzo},
  {Thilker}, {Annibali}, {Bressan}, \& {Buson}}]{Marino11b}
{Marino}, A., {Bianchi}, L., {Rampazzo}, R., {et~al.} 2011{\natexlab{a}}, \apj,
  736, 154

\bibitem[{{Marino} {et~al.}(2011{\natexlab{b}}){Marino}, {Rampazzo}, {Bianchi},
  {Annibali}, {Bressan}, {Buson}, {Clemens}, {Panuzzo}, \&
  {Zeilinger}}]{Marino11a}
{Marino}, A., {Rampazzo}, R., {Bianchi}, L., {et~al.} 2011{\natexlab{b}},
  \mnras, 411, 311

\bibitem[{{Marino} {et~al.}(2013){Marino}, {Plana}, {Rampazzo}, {Bianchi},
  {Rosado}, {Bettoni}, {Galletta}, {Mazzei}, {Buson}, {Ambrocio-Cruz}, \&
  {Gabbasov}}]{Ant13}
{Marino}, A., {Plana}, H., {Rampazzo}, R., {et~al.} 2013, \mnras, 428, 476

\bibitem[{{Martin} {et~al.}(2005){Martin}, {Fanson}, {Schiminovich}, \&
  et~al.}]{Martin05}
{Martin}, D.~C., {Fanson}, J., {Schiminovich}, D., \& et~al. 2005, \apjl, 619,
  L1

\bibitem[{{Martin} {et~al.}(2007){Martin}, {Wyder}, {Schiminovich}, {Barlow},
  {Forster}, {Friedman}, {Morrissey}, {Neff}, {Seibert}, {Small}, {Welsh},
  {Bianchi}, {Donas}, {Heckman}, {Lee}, {Madore}, {Milliard}, {Rich}, {Szalay},
  \& {Yi}}]{Martin07}
{Martin}, D.~C., {Wyder}, T.~K., {Schiminovich}, D., {et~al.} 2007, \apjs, 173,
  342

\bibitem[{{Mazzei}(2003)}]{paolaap}
{Mazzei}, P. 2003, Research Signapost, 37/661, India; Recent Research
  Developments in Astronomy \& Astrophysics, 1, 457

\bibitem[{{Mazzei}(2004)}]{paolaa}
---. 2004, arXiv:astro-ph/0401509

\bibitem[{{Mazzei} \& {Curir}(2003)}]{paola1}
{Mazzei}, P., \& {Curir}, A. 2003, \apj, 591, 784

\bibitem[{{Mazzei} \& {de Zotti}(1994)}]{paola6}
{Mazzei}, P., \& {de Zotti}, G. 1994, \apj, 426, 97

\bibitem[{{Mazzei} {et~al.}(1994){Mazzei}, {de Zotti}, \& {Xu}}]{paola4}
{Mazzei}, P., {de Zotti}, G., \& {Xu}, C. 1994, \apj, 422, 81

\bibitem[{{Mazzei} {et~al.}(2014){Mazzei}, {Marino}, {Rampazzo}, \& {et
  al}.}]{paolanew}
{Mazzei}, P., {Marino}, A., {Rampazzo}, R., \& {et al}. 2014,
  arXiv:astro-ph/1306.0777, Advanced in Space Research, Special Iusses:
  Ultraviolet Astrophysics, in press, 53

\bibitem[{{Mazzei} {et~al.}(1992){Mazzei}, {Xu}, \& {de Zotti}}]{paola2}
{Mazzei}, P., {Xu}, C., \& {de Zotti}, G. 1992, \aap, 256, 45

\bibitem[{{McDermid} {et~al.}(2006){McDermid}, {Emsellem}, {Shapiro}, {Bacon},
  {Bureau}, {Cappellari}, {Davies}, {de Zeeuw}, {Falc{\'o}n-Barroso},
  {Krajnovi{\'c}}, {Kuntschner}, {Peletier}, \& {Sarzi}}]{McD06}
{McDermid}, R.~M., {Emsellem}, E., {Shapiro}, K.~L., {et~al.} 2006, \mnras,
  373, 906

\bibitem[{{Navarro} {et~al.}(1997){Navarro}, {Frenk}, \& {White}}]{NFW97}
{Navarro}, J.~F., {Frenk}, C.~S., \& {White}, S.~D.~M. 1997, \apj, 490, 493

\bibitem[{{Nolan} {et~al.}(2007){Nolan}, {Dunlop}, {Panter}, {Jimenez},
  {Heavens}, \& {Smith}}]{Noetal06}
{Nolan}, L.~A., {Dunlop}, J.~S., {Panter}, B., {et~al.} 2007, \mnras, 375, 371

\bibitem[{{Paturel} {et~al.}(2003){Paturel}, {Petit}, {Prugniel}, {Theureau},
  {Rousseau}, {Brouty}, {Dubois}, \& {Cambr{\'e}sy}}]{Paturel03}
{Paturel}, G., {Petit}, C., {Prugniel}, P., {et~al.} 2003, \aap, 412, 45

\bibitem[{{Perea} {et~al.}(1990){Perea}, {del Olmo}, \& {Moles}}]{Peretal90}
{Perea}, J., {del Olmo}, A., \& {Moles}, M. 1990, \aap, 237, 319

\bibitem[{{Ramella} {et~al.}(2002){Ramella}, {Geller}, {Pisani}, \& {da
  Costa}}]{Ramella02}
{Ramella}, M., {Geller}, M.~J., {Pisani}, A., \& {da Costa}, L.~N. 2002, \aj,
  123, 2976

\bibitem[{{Rampazzo} {et~al.}(2013){Rampazzo}, {Panuzzo}, {Vega}, \& {et
  al.}.}]{Rampazzo13}
{Rampazzo}, R., {Panuzzo}, P., {Vega}, O., \& {et al.}. 2013, MNRAS~ accepted

\bibitem[{{Salim} {et~al.}(2012){Salim}, {Fang}, {Rich}, {Faber}, \&
  {Thilker}}]{Salimetal12}
{Salim}, S., {Fang}, J.~J., {Rich}, R.~M., {Faber}, S.~M., \& {Thilker}, D.~A.
  2012, \apj, 755, 105

\bibitem[{{Salim} \& {Rich}(2010)}]{Salim10}
{Salim}, S., \& {Rich}, R.~M. 2010, \apjl, 714, L290

\bibitem[{{Salim} {et~al.}(2005){Salim}, {Charlot}, {Rich}, {Kauffmann},
  {Heckman}, {Barlow}, {Bianchi}, {Byun}, {Donas}, {Forster}, {Friedman},
  {Jelinsky}, {Lee}, {Madore}, {Malina}, {Martin}, {Milliard}, {Morrissey},
  {Neff}, {Schiminovich}, {Seibert}, {Siegmund}, {Small}, {Szalay}, {Welsh}, \&
  {Wyder}}]{Salim05}
{Salim}, S., {Charlot}, S., {Rich}, R.~M., {et~al.} 2005, \apjl, 619, L39

\bibitem[{{Salpeter}(1955)}]{Salp55}
{Salpeter}, E.~E. 1955, \apj, 121, 161

\bibitem[{{Schawinski} {et~al.}(2007){Schawinski}, {Kaviraj}, {Khochfar},
  {Yoon}, {Yi}, {Deharveng}, {Boselli}, {Barlow}, {Conrow}, {Forster},
  {Friedman}, {Martin}, {Morrissey}, {Neff}, {Schiminovich}, {Seibert},
  {Small}, {Wyder}, {Bianchi}, {Donas}, {Heckman}, {Lee}, {Madore}, {Milliard},
  {Rich}, \& {Szalay}}]{Schaw07}
{Schawinski}, K., {Kaviraj}, S., {Khochfar}, S., {et~al.} 2007, \apjs, 173, 512

\bibitem[{{Sengupta} \& {Balasubramanyam}(2006)}]{SengeBal06}
{Sengupta}, C., \& {Balasubramanyam}, R. 2006, \mnras, 369, 360

\bibitem[{{Sil'chenko} {et~al.}(2010){Sil'chenko}, {Moiseev}, \&
  {Shulga}}]{Sietal10}
{Sil'chenko}, O.~K., {Moiseev}, A.~V., \& {Shulga}, A.~P. 2010, \aj, 140, 1462

\bibitem[{{Spavone} {et~al.}(2012){Spavone}, {Iodice}, {Bettoni}, {Galletta},
  {Mazzei}, \& {Reshetnikov}}]{marilena12}
{Spavone}, M., {Iodice}, E., {Bettoni}, D., {et~al.} 2012, \mnras, 426, 2003

\bibitem[{{Spavone} {et~al.}(2009){Spavone}, {Iodice}, {Calvi}, {Bettoni},
  {Galletta}, {Longo}, {Mazzei}, \& {Minervini}}]{marilena}
{Spavone}, M., {Iodice}, E., {Calvi}, R., {et~al.} 2009, \mnras, 393, 317

\bibitem[{{Strateva} {et~al.}(2001){Strateva}, {Ivezi{\'c}}, {Knapp},
  {Narayanan}, {Strauss}, {Gunn}, {Lupton}, {Schlegel}, {Bahcall}, {Brinkmann},
  {Brunner}, {Budav{\'a}ri}, {Csabai}, {Castander}, {Doi}, {Fukugita}, {Gy{\H
  o}ry}, {Hamabe}, {Hennessy}, {Ichikawa}, {Kunszt}, {Lamb}, {McKay},
  {Okamura}, {Racusin}, {Sekiguchi}, {Schneider}, {Shimasaku}, \&
  {York}}]{Strateva01}
{Strateva}, I., {Ivezi{\'c}}, {\v Z}., {Knapp}, G.~R., {et~al.} 2001, \aj, 122,
  1861

\bibitem[{{Temi} {et~al.}(2009){Temi}, {Brighenti}, \& {Mathews}}]{Temi09I}
{Temi}, P., {Brighenti}, F., \& {Mathews}, W.~G. 2009, \apj, 707, 890

\bibitem[{{Tinker} {et~al.}(2012){Tinker}, {George}, {Leauthaud}, {Bundy},
  {Finoguenov}, {Massey}, {Rhodes}, \& {Wechsler}}]{Tietal12}
{Tinker}, J.~L., {George}, M.~R., {Leauthaud}, A., {et~al.} 2012, \apjl, 755,
  L5

\bibitem[{{Trinchieri} {et~al.}(2012){Trinchieri}, {Marino}, {Mazzei},
  {Rampazzo}, \& {Wolter}}]{Trinetal12}
{Trinchieri}, G., {Marino}, A., {Mazzei}, P., {Rampazzo}, R., \& {Wolter}, A.
  2012, \aap, 545, A140

\bibitem[{{Valdarnini}(2002)}]{Val2002}
{Valdarnini}, R. 2002, \apj, 567, 741

\bibitem[{{Villalobos} {et~al.}(2012){Villalobos}, {.}, {De Lucia}, {Borgani},
  \& {Murante}}]{Villalobos12}
{Villalobos}, {\'A}., {.}, {De Lucia}, G., {Borgani}, S., \& {Murante}, G.
  2012, \mnras, 424, 2401

\bibitem[{{Warren} {et~al.}(1992){Warren}, {Quinn}, {Salmon}, \&
  {Zurek}}]{war92}
{Warren}, M.~S., {Quinn}, P.~J., {Salmon}, J.~K., \& {Zurek}, W.~H. 1992, \apj,
  399, 405

\bibitem[{{Wei} {et~al.}(2010){Wei}, {Kannappan}, {Vogel}, \&
  {Baker}}]{Wiletal10}
{Wei}, L.~H., {Kannappan}, S.~J., {Vogel}, S.~N., \& {Baker}, A.~J. 2010, \apj,
  708, 841

\bibitem[{{Welch} {et~al.}(2010){Welch}, {Sage}, \& {Young}}]{Weletal10}
{Welch}, G.~A., {Sage}, L.~J., \& {Young}, L.~M. 2010, \apj, 725, 100

\bibitem[{{Wetzel} {et~al.}(2012){Wetzel}, {Tinker}, \&
  {Conroy}}]{Wetzeletal12}
{Wetzel}, A.~R., {Tinker}, J.~L., \& {Conroy}, C. 2012, \mnras, 424, 232

\bibitem[{{Wilman} {et~al.}(2009){Wilman}, {Oemler}, {Mulchaey}, {McGee},
  {Balogh}, \& {Bower}}]{Wilman09}
{Wilman}, D.~J., {Oemler}, Jr., A., {Mulchaey}, J.~S., {et~al.} 2009, \apj,
  692, 298

\bibitem[{{Wyder} {et~al.}(2007){Wyder}, {Martin}, {Schiminovich}, {Seibert},
  {Budav{\'a}ri}, \& {Treyer}}]{Wyderetal07}
{Wyder}, T.~K., {Martin}, D.~C., {Schiminovich}, D., {et~al.} 2007, \apjs, 173,
  293

\bibitem[{{Yamamura} {et~al.}(2009){Yamamura}, {Makiuti}, {Ikeda}, {Fukuda},
  {Yamauchi}, {Hasegawa}, {Nakagawa}, {Narumi}, {Baba}, {Takagi}, {Jeong},
  {Oh}, {Lee}, {Savage}, {Rahman}, {Thomson}, {Oliver}, {Figueredo},
  {Serjeant}, {White}, {Pearson}, {Wang}, {Rowan-Robinson}, {Kester}, {van der
  Wolk}, {Barthel}, {Salama}, {Alfageme}, {Garc{\'{\i}}a-Lario}, {Stephenson},
  {Cohen}, \& {Mueller}}]{Yamaetal09}
{Yamamura}, I., {Makiuti}, S., {Ikeda}, N., {et~al.} 2009, in Astronomical
  Society of the Pacific Conference Series, Vol. 418, AKARI, a Light to
  Illuminate the Misty Universe, ed. T.~{Onaka}, G.~J. {White}, T.~{Nakagawa},
  \& I.~{Yamamura}, 3

\bibitem[{{Young} {et~al.}(2011){Young}, {Bureau}, {Davis}, {Combes},
  {McDermid}, {Alatalo}, {Blitz}, {Bois}, {Bournaud}, {Cappellari}, {Davies},
  {de Zeeuw}, {Emsellem}, {Khochfar}, {Krajnovi{\'c}}, {Kuntschner},
  {Lablanche}, {Morganti}, {Naab}, {Oosterloo}, {Sarzi}, {Scott}, {Serra}, \&
  {Weijmans}}]{Youngetal11}
{Young}, L.~M., {Bureau}, M., {Davis}, T.~A., {et~al.} 2011, \mnras, 414, 940

\end{thebibliography}

 \end{document}